\documentclass[a4paper,11pt]{article}

\usepackage{jheppub}
\allowdisplaybreaks
\usepackage{graphicx}
\usepackage{epstopdf}

\newcommand{\be}{\begin{equation}}
\newcommand{\ee}{\end{equation}}
\newcommand{\bea}{\begin{eqnarray}}
\newcommand{\eea}{\end{eqnarray}}
\newcommand{\nn}{\nonumber}

\title{EFTofPNG: \\
A package for high precision computation with the \\
Effective Field Theory of Post-Newtonian Gravity}

\author[a]{Michele Levi}
\author[b]{and Jan Steinhoff}

\affiliation[a]{Institut de Physique Th\'eorique, Universit\'e Paris-Saclay, CEA, CNRS,\\ 
91191 Gif-sur-Yvette, France}

\affiliation[b]{Max-Planck-Institute for Gravitational Physics 
(Albert-Einstein-Institute),\\ Am M{\"u}hlenberg 1, 14476 Potsdam-Golm, Germany}

\emailAdd{michele.levi@ipht.fr}
\emailAdd{jan.steinhoff@aei.mpg.de}

\abstract{We present a novel public package ``EFTofPNG'' for high precision computation in 
the effective field theory of post-Newtonian (PN) Gravity, including spins. We created this 
package in view of the timely need to publicly share automated computation tools, which 
integrate the various types of physics manifested in the expected increasing influx of 
gravitational wave (GW) data. Hence, we created a free and open source package, which is 
self-contained, modular, all-inclusive, and accessible to the classical Gravity community. 
The ``EFTofPNG'' Mathematica package also uses the power of the ``xTensor'' package, suited 
for complicate tensor computation, where our coding also strategically approaches the 
generic generation of Feynman contractions, which is universal to all perturbation theories 
in physics, by efficiently treating n-point functions as tensors of rank n. The package 
currently contains four independent units, which serve as subsidiaries to the main one. Its 
final unit serves as a pipeline chain for the obtainment of the final GW templates, and 
provides the full computation of derivatives and physical observables of 
interest. The upcoming ``EFTofPNG'' package version 1.0 should cover the point mass sector, 
and all the spin sectors, up to the fourth PN order, and the two-loop level. We expect and 
strongly encourage public development of the package to improve its efficiency, and to 
extend it to further PN sectors, and observables useful for the waveform modelling.}

\begin{document}

\maketitle

\flushbottom

\section{Introduction}
\label{intro}

The recent direct detection of gravitational wave (GW) events GW150914 \cite{Abbott:2016blz} 
and GW151226 \cite{Abbott:2016nmj} from binary black hole (BH) coalescence herald a new era 
of GW astronomy. With the Advanced Virgo detector in Europe \cite{Virgo} to join the second 
observational run of the twin Advanced LIGO detectors in the US \cite{LIGO}, and an improved 
signal sensitivity, even more upcoming detections are expected, including binaries that also 
consist of neutron stars (NS). Yet, in order to go beyond merely detecting more GW events, 
and truly enter an era of high precision Gravity, accurate theoretical waveforms are
required. These waveforms should integrate the various physical phases, and the various 
corresponding physical approaches to model the continuous signal, which is identified with 
the matched filtering technique. The initial part of the waveform corresponds to the 
inspiral phase of the binaries' evolution, where the orbital velocity is non-relativistic, 
and can only be described analytically, via the post-Newtonian (PN) approximation of General 
Relativity (GR) \cite{Blanchet:2013haa, TheLIGOScientific:2016src}. Indeed, though 
numerical simulations can handle the merger and ringdown phases of the coalescence, the 
typical long time scale in the inspiral phase makes it unrealistic to tackle it numerically.

In fact, high order PN corrections, in particular also ones, which take into account the 
\textit{spins} of the compact objects, are required to be incorporated into the 
Effective-One-Body (EOB) \cite{Buonanno:1998gg} modelling of the waveforms. This would allow 
to improve the analysis of the GW events, and hence gain more information on the intrinsic 
parameters of the compact objects \cite{TheLIGOScientific:2016wfe, Abbott:2016izl}. This 
additional information has implications for our knowledge of the astrophysics of such 
binaries, the physics of compact objects, and notably of the fundamental theory of Gravity. 
Actually, the remarkably high fourth PN (4PN) order correction at the point mass sector has 
been recently tackled and completed by various methods in multiple works 
\cite{Jaranowski:2012eb,Foffa:2012rn,Jaranowski:2013lca,Damour:2014jta,Damour:2016abl,Bernard:2015njp,Bernard:2016wrg,Damour:2017ced,Foffa:2016rgu}. 
Simultaneously, a recent series of works 
\cite{Levi:2011eq,Levi:2014sba,Levi:2014gsa,Levi:2015msa,Levi:2015uxa,Levi:2015ixa,Levi:2016ofk}, 
based on an effective field theory (EFT) approach for the binary inspiral 
\cite{Goldberger:2004jt,Goldberger:2007hy}, has accomplished similar PN accuracy for the 
more intricate spin sectors, which account for generic compact spinning objects in the 
binaries. This line of work includes all spin interactions linear and quadratic in spin up 
to the next-to-next-to-leading order (NNLO), and at the leading order (LO) cubic and quartic 
in spin, which were obtained via an EFT for spinning objects, formulated in 
\cite{Levi:2015msa} (we note a line of work, also originating from the EFT approach in 
\cite{Goldberger:2004jt,Goldberger:2007hy}, that first tackled spinning objects for this 
sector in \cite{Porto:2005ac}, and up to next-to-leading order (NLO) in 
\cite{Porto:2008tb,Porto:2008jj}).

In this paper we introduce a novel public package ``EFTofPNG'' for high precision 
computation in the EFT of PN Gravity, including spins. We created this package with the view 
of the close analogy between the impending progress in Gravity, and that which occurred in, 
e.g., Cosmology, where in the past few decades, there has been a tremendous improvement in 
the quality and quantity of observations and experiments. In Cosmology this has led to the 
imperative development of comprehensive public computation software, like CAMB 
\cite{Lewis:1999bs,CAMB}, which has become regularly used by the whole of the community. 
Similarly to Cosmology, we need to share automated computation tools, which could then be 
assembled to even more inclusive public codes, that integrate the various types of 
physics manifested in the increasing influx of GW data, which will soon become public, 
e.g.~through the LIGO Open Science Center \cite{LigoLosc}. 

In the Gravity community there are already public codes, notably, e.g., the Einstein Toolkit 
\cite{EinsTK} for numerical relativity. Moreover, in the GW community there are the 
extensive Ligo's LALSuite \cite{LigoDA}, which is comprised of various GW data analysis 
components, such as LALSimulation with waveform models, including EOB and phenomenological 
models, and the detection pipelines GstLAL \cite{LigoDA} and PyCBC \cite{pycbc}. The goal of 
our current code is to start filling in the gap in the analytic part of the waveform models, 
going from the ``full theory'' in the inspiral phase via a tower of EFTs to the PN 
observables that enter waveform models in LALSimulation, namely to complete the current GW 
codes, and to foster further collaboration in the community.

The EFT approach is naturally tailored for high precision computation, but 
previous automatic PN computations in the EFT approach made use of patches of less relevant 
public codes, such as ``FeynCalc'' from particle physics \cite{Chu:2008xm,Foffa:2011ub}. In 
addition, these automated computations were limited to the point mass sector, and most were 
not made public. Hence, our goal was to create a free and open source code, which is 
self-contained, modular, all-inclusive, and accessible to the Gravity, and particularly to 
the GW community. In fact, this package may benefit a broader community, as, 
e.g., our coding strategically approaches the generic generation of Feynman contractions and 
graphs, which is universal to all perturbation theories in physics, by efficiently treating 
$n$-point functions as tensors of rank $n$.

This paper is organized as follows.
In section \ref{theory} we start by reviewing the formulation of the Effective Field Theory 
of PN Gravity, including spinning objects. In section \ref{overview} we present an overview 
of the EFTofPNG package version 1.0, which currently contains four independent units, that 
serve as subsidiaries to its main one. We then proceed to go over the different units of the 
package. In section \ref{main} we present the ``Main'' unit, where the complete evaluation 
of the Feynman diagrams of the desirable sectors is done. In section \ref{feynrul} we 
describe the ``FeynRul'' unit, which evaluates the Feynman rules, and should be run first to 
provide input to other units. Next, in section \ref{feyngen} we describe the ``FeynGen'' 
unit, which generates the required Feynman contractions and diagrams needed to be evaluated 
by the ``Main'' unit. In section \ref{nloop} we present the ``NLoop'' unit, which 
independently produces all the required loop master integrals with their simplified forms to 
be output to the ``Main'' unit. Lastly, in section \ref{giobs} we note the useful 
independent and final unit of the package, ``GI observables'', which using the output of the 
``Main'' unit, provides derivatives and gauge invariant (GI) observables of interest, and 
serves as a pipeline chain for the obtainment of the final GW templates for the detectors. 
Finally, in section \ref{theendmyfriend} we present our main conclusions.

\section{The EFT of PNG including spins}
\label{theory}

The EFT of PN Gravity for extended objects, which we build on, has been introduced in 
\cite{Goldberger:2004jt,Goldberger:2007hy}. The basic idea is to make use of the natural 
hierarchy of scales, that exists in the binary inspiral problem, and thus tackle it by 
constructing a tower of EFTs for each of the scales. It holds that $r_s\sim r\,v^2 \sim 
\lambda \, v^3$, where $r_s$ is the scale of the internal structure of the single compact 
component of the binary, $r$ is the orbital separation scale, $\lambda$ is the radiation 
wavelength scale, and $v$ is the typical non-relativistic orbital velocity at the inspiral 
phase, i.e.~$v\ll 1$. We also have that $r_s\sim m$, where $m$ is the mass of the isolated 
compact object, and is the only scale in the full theory. Yet we note that we consider in 
our work the generic case of \textit{spinning} gravitating objects. While a spinning point 
particle is also characterized by its spin length, $S^2$,  it holds that $S\lesssim m^2$ 
\cite{Levi:2015msa}.

For each EFT in the tower the field is decomposed to its Fourier modes, which can be 
assigned a definite power counting in terms of the small PN parameter, $v$. First, the EFT 
for an isolated compact object is constructed with an effective action of the generic 
following form:
\be\label{seffsingobj}
S_{\text{eff}}\left[\bar{g}_{\mu\nu},y^\mu,e^{\mu}_{A}\right]=
-\frac{1}{16\pi G} \int d^4x \sqrt{\bar{g}} R\left[\bar{g}_{\mu\nu}\right] + 
\underbrace{\sum_{i}C_i\int d\sigma O_i(\sigma)}_{S_{pp}\equiv
\text{point particle action}},
\ee 
where an infinite tower of worldline operators, $O_i(\sigma)$, consistent with the 
symmetries of the theory, is introduced. Thus, all UV dependence shows up only in the Wilson 
coefficients, $C_i(r_s)$, in the point particle action, $S_{\text{pp}}$. Here, 
$\bar{g}_{\mu\nu}$ represents the field modes far above the single object scale, and 
$y^\mu$, $e^{\mu}_{A}$, are the particle worldline coordinate, and worldline tetrad degrees 
of freedom (DOFs), respectively. We discuss in detail the relevant symmetries and DOFs for 
the general case of spinning gravitating objects in \cite{Levi:2015msa}.

Next, for the EFT of the binary, which is then to be regarded as a single composite object, 
we first write the decomposition of the field modes as
$\bar{g}_{\mu\nu}\equiv\eta_{\mu\nu}+H_{\mu\nu}+\widetilde{h}_{\mu\nu}$.
Starting from the effective action of two compact objects, with a copy of the point particle 
action for each of them, the effective action of the composite object is then
obtained by explicitly integrating out the field modes below the orbital scale, 
$H_{\mu\nu}$, namely the effective action is defined by the following functional integral:
\be \label{orbitintegrateout}
e^{iS_{\text{eff(composite)}}\left[\widetilde{h}_{\mu\nu}, y^\mu, e^{\mu}_A\right]} 
\equiv \int {\cal{D}}H_{\mu\nu}~e^{iS_{\text{eff}}\left[\bar{g}_{\mu\nu}, y^\mu_{1}, 
y^\mu_{2}, e_{(1)}{}^{\mu}_{A}, e_{(2)}{}^{\mu}_{A}\right]}, 
\ee
considering its classical limit. Here $y^\mu$ and $e^{\mu}_A$ on the left hand side are the 
worldline coordinate and tetrad, respectively, of the single composite particle.

To obtain an EFT of radiation, the final EFT in the tower, one should proceed in general to 
also integrate out the radiation modes of the gravitational field, $\widetilde{h}_{\mu\nu}$. 
Yet, in the conservative sector, where no radiation modes are present, no further process is 
required after the field modes below the orbital scale have been integrated out. Hence, we 
emphasize that the final effective action in this sector, should be one without any 
remaining orbital scale field DOFs. This implies that all unphysical DOFs should also be 
eliminated from the action, in particular those associated with the rotational DOFs 
\cite{Levi:2008nh,Levi:2014sba,Levi:2015msa}. 

For the point particle action in eq.~\eqref{seffsingobj} of a spinning object, we start from 
the form \cite{Hanson:1974qy,Bailey:1975fe,Levi:2015msa}:
\begin{align}\label{Spp}
S_{\text{pp}} = \int d \sigma \left[ -m \sqrt{u^2}
								- \frac{1}{2} S_{\mu\nu}\Omega^{\mu\nu}
                + L_{\text{SI}}\left[u^{\mu}, S_{\mu\nu}, \bar{g}_{\mu\nu}
                \left(y^\mu\right)\right]\right],
\end{align} 
where $u^\mu$ is the coordinate velocity, $\Omega^{\mu\nu}$, the angular velocity, 
$S_{\mu\nu}$, the spin, conjugate to the angular velocity, and $L_{\text{SI}}$ stands for 
the nonminimal coupling part of the action, that consists only of the spin-induced 
multipoles. In \cite{Levi:2015msa} we worked out both the minimal coupling and nonminimal 
coupling parts of the point particle action in order to obtain the effective action of a 
single spinning particle, i.e.~the first EFT in the tower that we noted. 

First, we restore the gauge freedom of the rotational variables by performing a boost-like 
transformation in a covariant form on the worldline tetrad \cite{Levi:2015msa}. This 
leads to a generic gauge for the tetrad, and the spin, which reads:
\begin{equation}\label{gengauge+SSC}
\hat{e}_{[0]\mu} = w_{\mu}, \qquad
\hat{S}^{\mu\nu} \left( p_{\nu} + \sqrt{p^2} \hat{e}_{[0]\nu} \right) = 0,
\end{equation}
where a new gauge DOF, $w_{\mu}$, is introduced as the timelike vector of the tetrad. All in 
all, this provides the $3+3$ necessary gauge constraints to eliminate the redundant 
unphysical DOFs in the 4D antisymmetric angular velocity and spin tensors. For the minimal 
coupling term we then obtain:
\begin{align}\label{mcTrans}
\frac{1}{2} S_{\mu\nu} \Omega^{\mu\nu} &= 
  \frac{1}{2} \hat{S}_{\mu\nu} \hat{\Omega}^{\mu\nu}
	+ \frac{\hat{S}^{\mu\nu} p_{\nu}}{p^2} \frac{D p_{\mu}}{D \sigma},
\end{align}
thus an extra term appears in the action, which contributes to finite size effects, yet 
carries no Wilson coefficients. As for the nonminimal coupling spin-induced part of the 
action, since as we argue it only depends on the spin, one only needs to transform to the 
generic spin variable with 
\begin{equation}\label{Strans}
S_{\mu\nu} = \hat{S}_{\mu\nu} - \frac{\hat{S}_{\mu\rho} p^{\rho} p_{\nu}}{p^2}
	+ \frac{\hat{S}_{\nu\rho} p^{\rho} p_{\mu}}{p^2}.
\end{equation} 

Moreover, the nonminimal coupling spin-induced part of the point particle action should be 
constrained. This is indeed accomplished using the symmetries of the theory, and the various 
considerations in the problem, as described in detail in \cite{Levi:2015msa}. Hence, we find 
that the LO nonminimal couplings to all orders in spin are fixed as follows: 
\begin{align} \label{sinmc}
L_{\text{SI}}=&\sum_{n=1}^{\infty} \frac{\left(-1\right)^n}{\left(2n\right)!}
\frac{C_{ES^{2n}}}{m^{2n-1}} D_{\mu_{2n}}\cdots D_{\mu_3}
\frac{E_{\mu_1\mu_2}}{\sqrt{u^2}} S^{\mu_1}S^{\mu_2}\cdots 
S^{\mu_{2n-1}}S^{\mu_{2n}}\nn\\
&+\sum_{n=1}^{\infty} \frac{\left(-1\right)^n}{\left(2n+1\right)!}
\frac{C_{BS^{2n+1}}}{m^{2n}} 
D_{\mu_{2n+1}}\cdots D_{\mu_3}\frac{B_{\mu_1\mu_2}}{\sqrt{u^2}} 
S^{\mu_1}S^{\mu_2}\cdots 
S^{\mu_{2n-1}}S^{\mu_{2n}}S^{\mu_{2n+1}},
\end{align}
where we have introduced new spin-induced Wilson coefficients, and the new operators consist 
of the electric and magnetic curvature tensors, $E_{\mu\nu},$ and $B_{\mu\nu},$ 
respectively, and the spin vector $S^{\mu}$, as their building blocks. In particular up to 
the 4PN order, the quadrupole, octupole, and hexadecapole couplings are required 
\cite{Levi:2014gsa,Levi:2015msa,Levi:2015ixa}, which explicitly read:
\begin{align} \label{es2bs3es4}
L_{ES^2} =& -\frac{C_{ES^2}}{2m} \frac{E_{\mu\nu}}{\sqrt{u^2}} 
S^{\mu} S^{\nu},\\
L_{BS^3}=&-\frac{C_{BS^3}}{6m^2}\frac{D_\lambda B_{\mu\nu}}
{\sqrt{u^2}}S^{\mu} S^{\nu} S^{\lambda},\\
L_{ES^4}=&\frac{C_{ES^4}}{24m^3} \frac{D_\lambda D_\kappa 
E_{\mu\nu}}{\sqrt{u^2}} S^{\mu} S^{\nu} S^{\lambda} S^{\kappa}.
\end{align}

Since we work in an action approach the gauge of the rotational variables should be fixed at 
the level of the point particle action as was put forward in \cite{Levi:2008nh}. In 
particular, in order to obtain the next EFT, the particle DOFs should be disentangled from 
the field DOFs, and this can only be obtained once the gauge for the rotational variables is 
fixed. Then the minimal coupling term from eq.~\eqref{mcTrans} can be written as 
\cite{Levi:2010zu,Levi:2015msa}:
\begin{align} \label{Sfieldsplit}
\frac{1}{2} \hat{S}_{\mu\nu} \hat{\Omega}^{\mu\nu}
&= \frac{1}{2} \hat{S}_{ab} \hat{\Omega}^{ab}_{\text{flat}}
         + \frac{1}{2} \hat{S}_{ab} \omega_{\mu}{}^{ab} u^{\mu},      
\end{align}
where we have used the Ricci rotation coefficients, defined with the tetrad field, 
$\tilde{e}^{a\nu}$, by
$\omega_{\mu}{}^{ab} \equiv \tilde{e}^b{}_{\nu} D_{\mu}\tilde{e}^{a\nu}$,
 and we switched to new rotational variables: $\hat{\Omega}^{ab}_{\text{flat}} 
\equiv\hat{\Lambda}^{Aa} \frac{d \hat{\Lambda}_A{}^b}{d \sigma}$, the locally 
flat spacetime angular velocity tensor with the worldline Lorentz matrices, 
$\hat{\Lambda}_A{}^a$, and 
$\hat{S}_{ab}\equiv\tilde{e}^{\mu}_{a}\tilde{e}^{\nu}_{b}\hat{S}_{\mu\nu}$, the spin 
projected to the local frame. Before we integrate out the orbital scale we need to fix all 
the gauges. Before that, we apply the beneficial non-relativistic space+time decomposition 
of the gravitational field, i.e.~the non-relativistic gravitational (NRG) fields 
\cite{Kol:2007bc, Kol:2010ze}. The tetrad field gauge is then fixed to the time gauge of 
Schwinger, that corresponds to the NRG parametrization, and we fix the gauge of the 
rotational variables to that we dub ``the canonical gauge''. 

Finally, once the effective action of the binary object is obtained, the equations of 
motion (EOMs) of both the positions, and of the spins, are obtained directly via a proper 
variation of the action. In particular for the EOMs of spin an independent variation should 
be made with respect to the spin and to its conjugates, the Lorentz matrices, as explained 
in \cite{Levi:2014sba}. Then a simple form for the EOMs of spin is obtained, which contains 
only physical DOFs. In addition, it is straightforward to obtain the Hamiltonians in the 
standard manner as in the non-spinning case.

\section{The EFTofPNG package version 1.0}
\label{overview}
 
The main purpose of creating ``EFTofPNG'' as free and open source software is twofold: 1.~On 
the one hand we want to make the EFT of PNG accessible in its most practical form to the 
classical Gravity community, who wish to understand the EFT of PNG better, and possibly 
build on it, and extend it; 2.~On the other hand for those in the Gravity, and in particular 
the GW community, who only wish to use the outcome of implementing the EFT of PNG in order 
to further process its results into actual models of gravitational waveforms, we want to 
provide the complete pipeline, which contains the full computation of derivatives and gauge 
invariant observables of interest. 

The ``EFTofPNG'' package is written in Wolfram language of Mathematica, and requires in 
addition the ``xTensor'' package for abstract tensor algebra from the ``xAct'' bundle 
\cite{xAct}. The time is ripe for the community to have a self-contained package that applies 
Feynman calculus specifically in the novel context of classical Gravity. Rather than rely on 
patches of less efficient and suitable existing packages from the conventional context of 
particle physics, this package uses the power of the ``xTensor'' package, naturally suited 
for the complicate tensor computation required in Gravity, and which is familiar to the 
Gravity community. Moreover, our package strategically approaches the generic generation of 
Feynman contractions and graphs, which is universal to all perturbation theories, classical 
or quantum, in all fields of physics, by regarding $n$-point functions as tensors of rank 
$n$, as further explained in section \ref{feyngen} below. Hence our package also benefits in 
this broad respect from the full power of the ``xTensor'' package, which is worth 
disseminating also outside of the Gravity community. 

\begin{figure}[t]
\includegraphics[width=\linewidth]{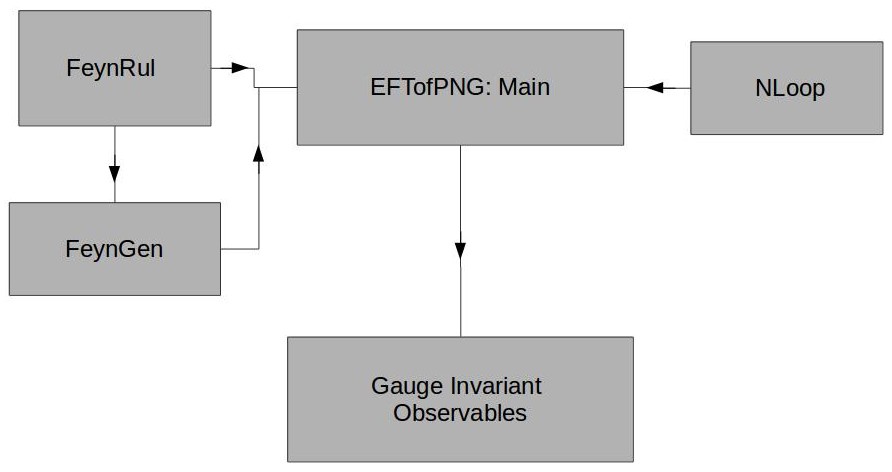}
\caption{Outline of the ``EFTofPNG'' package version 1.0. Details on the different units of 
the package, and the dependencies among them are given below.}\label{packeftofpng}
\end{figure}

In figure \ref{packeftofpng} a schematic outline of the ``EFTofPNG'' package version 1.0 is 
shown. The package currently contains four independent units, which serve as subsidiary ones 
to the ``Main'' unit of the package. First, the ``FeynRul'' unit, which evaluates the 
Feynman rules, should be run in order to provide input both to the ``FeynGen'' unit, and 
directly to the ``Main'' unit. Next, the ``FeynGen'' unit should be run to generate the 
required Feynman contractions and diagrams needed to be evaluated by the ``Main'' unit. In 
parallel, the ``NLoop'' unit, which produces all the required loop master integrals, and 
their most simplified forms, should be run, to also provide direct input to the ``Main'' 
unit. Finally, after the complete evaluation of the Feynman diagrams is done by running the 
``Main'' unit, its output is supplied to the independent ``GI observables'' unit, which 
provides derivatives and gauge invariant physical observables of interest, and serves as 
a pipeline chain for the obtainment of the final GW templates for the detectors. In the 
sections that follow we elaborate on each of the abovementioned units of the package. 

All of the units in the package use a generic $d+1$ dimensional space-time setting,  
since all of the Feynman integrals are evaluated using dimensional regularization.
The PN order parameter in the package is the standard inverse of c, the speed of light, 
denoted as ``cInv'' all through the package. It should also be noted that the PN power 
counting is the standard PN count, where the Newtonian interaction is assigned the 
$(\text{cInv})^0$ coefficient in the various series expansions. 

A crucial aspect for the efficiency of the computation is that the NRG fields are applied 
all through the package. Whereas for a general field parametrization
the N$^{\text{n}}$LO, that is the (next-to)$^{\text{n}}$-leading order, requires a 
computation at the $n$-loop level, the use of the NRG fields entails several 
simplifications, in particular that in the point mass sectors the N$^{\text{n}}$LO requires 
only the $2\lfloor n/2\rfloor$ loop level, such that e.g., the 3PN order in the point mass 
sector, which is NNNLO$\equiv$N$^3$LO in this sector, requires only a two-loop computation 
\cite{Kol:2007bc, Kol:2010ze}, rather than a three-loop one. In the spin sectors the 
N$^\text{n}$LO still corresponds to a computation at the $n$-loop level, but fortunately in 
these sectors the N$^\text{n}$LO is shifted to at least the ($n+1.5$)PN order. 

The upcoming ``EFTofPNG'' package version 1.0 is expected to cover the point mass sector up 
to the 3PN order, and all the spin sectors up to the 4PN order, including the NNLO up to 
quadratic in the spins \cite{Levi:2011eq,Levi:2015uxa,Levi:2015ixa}, and the LO cubic and 
quartic in the spins \cite{Levi:2014gsa}. The current preliminary version 0.99 is already 
released as a public repository in GitHub, and can be found in the URL: 
``https://github.com/miche-levi/pncbc-eftofpng''. 
We expect and strongly encourage further public development of the package to improve its 
efficiency and run times, and to further extend it to higher PN and spin orders, to 
non-conservative sectors, and to further observables useful for the waveform modeling, for 
the benefit of the community. A file, which details the coding style conventions to be 
followed by future developers, is included in the public repository, in order to keep the 
package as accessible and readable as possible. 
The repository contains a README file, and the code itself contains specific documentation, 
and different examples of actual computations. 
Moreover, results of various runs are also provided in the repository for those who wish to only make use of some final outputs. 
For the complete details of the package 
please refer to the abovementioned public repository.

\subsection{Main: Evaluating the two-body effective action}
\label{main}

The ``Main'' unit of the ``EFTofPNG'' package is where the evaluation of the effective 
action of the binary eventually takes places. This unit consists of independent interfaces 
with the outputs of the ``FeynRul'' and ``FeynGen'' units, followed by a ``FeynInteg'' 
section, which performs all possible processing required for the evaluation of the Feynman 
diagrams, using also the output of the ``NLoop'' unit for the loop master integrals. The 
unit then ends with a ``FeynComp'' section, which carries out an automatic routine that 
evaluates altogether all the Feynman diagrams for the specified desirable sectors, while 
also demonstrating some singular evaluations of more complex diagrams by ``manually'' 
specifying the corresponding contractions.

Let us then go through the sections of the ``Main'' unit in more detail.
The unit starts with a general ``Setup'' section. Initial definitions are made within 
xTensor of a $d$-dimensional flat manifold, and the time coordinate as the worldline 
parameter. All worldline DOFs, e.g.~the positions and spins, are defined as tensors on this 
flat manifold. Next, heads are defined to label the worldline and general spacetime points, 
as well as the time derivatives of these tensors. In addition, a manifold for constant 
tensors is defined for, e.g., the Wilson coefficients (these are not treated as dynamic DOFs 
in the context of this work). Finally, generic tools for the ``Main'' unit are set up, and 
order parameters for the PN count, and the spin, are defined in order to distinguish among 
the various PN, as well as the point mass and spin, sectors. The subsequent sections in 
``Main'' are described below.
 
\paragraph{FeynRul: Preparing the Feynman rules.}

At the next stage the ``FeynRul'' section complements and processes the Feynman rules 
obtained as an output of the independent ``FeynRul'' unit, presented in section 
\ref{feynrul} below. First, the NRG fields, and the Dirac delta of time, are defined on the 
general manifold. The Fourier vectors are defined on the manifold for constant tensors. 
Further heads are defined for the differentials of integrals, the general Dirac delta, and 
the propagators scalar. All the bulk and worldline vertices are then defined on a new 
``vertices'' manifold as further explained in section \ref{feyngen} below. 

Next, the Feynman rules are imported from the output files of the ``FeynRul'' unit, to be 
followed by their processing. They are transformed to Fourier space, including momentum 
conservation in the bulk vertices. The time dependence of the latter is handled, such that 
it finally drops on the worldline vertices. The propagator and propagator time insertions 
are then defined, and the bulk and worldline vertices uniquely labeled.

\paragraph{FeynGen: Processing the Feynman diagrams.}

The subsequent ``FeynGen'' section imports the Feynman diagrams generated in the  
independent ``FeynGen'' unit, presented in section \ref{feyngen} below. The various 
components of the imported diagrams are then converted to conform with the corresponding  
components as defined within the ``Main'' unit.

\paragraph{FeynInteg: Handling the Feynman integration.}

The ``FeynInteg'' section constitutes the core of the ``Main'' unit. This section tells how 
to manipulate the Feynman integrals that should be evaluated. First, the time derivative, 
and how to handle it with partial integrations, are defined. All the integrals, and 
specifically also the time integrals, are handled using distinctive heads. Dirac deltas of 
momenta are integrated over with a simplification of the intermediate propagators. Multiple 
integrals are factorized, and the loop integrals are nested. Intermediate required 
transformations of integration variables are handled. Then, the loop master integrals, 
needed in order to actually evaluate the Feynman integrals, are imported from the output 
file of the independent ``NLoop'' unit, presented in section \ref{nloop} below. 

Finally, an ``IntegrateF'' function is defined, where all the previous operations are 
applied, and the regularization limit in the dimension is taken. This function can be 
applied on individual specified contractions. In addition, an automatic integration function 
is defined, which operates on sums of given contractions. We demonstrate the use of both 
functions in the final evaluation of the two-body effective action, done in the ``FeynComp'' 
section that follows.   

\paragraph{FeynComp: Evaluating the Feynman diagrams.}

In the final ``FeynComp'' section the evaluation of various PN orders, point mass and spin 
sectors, is demonstrated. Feynman integration can be performed either collectively with the 
automatic integration function described above, or specifically for individual Feynman 
diagrams, where the contraction is ``manually'' specified. Indeed, the evaluation of a few 
irreducible two-loop diagrams is demonstrated there. Finally, the diagram contributions are 
summed to yield the corresponding interaction parts of the two-body effective action, and
the total effective action is exported as the output of the ``Main'' unit. The overall run 
time of the ``Main'' unit applied to all sectors up to the two-loop level, and quadratic in 
the spins is currently of about five days.

\subsection{FeynRul: Deriving the Feynman rules}
\label{feynrul}

The ``FeynRul'' unit of the package takes the first step from the theory presented in 
section \ref{theory}. It actually evaluates the effective action of the single spinning 
particle, while fixing the necessary gauges, and derives the Feynman rules, that are later 
used to obtain the EFT of the binary including spins. 

In this unit we define space+time decomposed components of the metric, inverse metric, and 
mixed tetrads, over a flat background $d$-dimensional manifold, where d is the generic 
number of spatial dimensions. We gauge the worldline parameter to be the time coordinate, 
and we define the worldline DOFs as tensors on the flat manifold. For this reason we make 
our own $d+1$ definitions of the Christoffel symbol and the Riemann curvature independent of 
xTensor, and use $d+1$ objects, which are then embedded into the $d$-dimensional xTensor 
manifold. As we noted in section \ref{overview} we use the NRG fields, with the 
corresponding Schwinger time gauge for the tetrads.   

The bulk gravitational action is computed using the Einstein-Hilbert form, but other 
equivalent forms, e.g.~the Gamma-Gamma form, differing by a total divergence, can be used. 
Indeed, we also demonstrate an addition of a total divergence in the code in order to 
conform with the specific form of previous results, which were presented in, 
e.g., \cite{Levi:2015ixa,Levi:2015uxa,Levi:2016ofk}. The harmonic gauge is used for the 
gravitational field. Up to the PN orders currently evaluated, and using the NRG fields, we 
need up to quartic self-interaction in the bulk action, i.e.~up to four-point functions.
The propagators, and the quadratic time insertions, regarded as perturbative corrections to 
the propagators, are inserted separately. 
 
As described in section \ref{theory}, in the spin part of the minimal coupling action there 
is an extra term, coming from rotational gauge freedom, and then the gauge is fixed to the 
canonical gauge. In the nonminimal coupling spin-induced action, which involves the Riemann 
dependent operators, we begin by considering the covariant spin, and the automatic 
evaluation becomes more time-consuming. The run time required for the bulk action, and for 
the mass and linear-in-spin parts of the point particle action is quite short (circa few 
minutes), but becomes longer with higher orders of the spin in the nonminimal coupling part 
(about ten to twenty minutes). Finally, the bulk vertices, and worldline vertices, are 
output in two separate files, which are used independently by the ``FeynGen'', and ``Main'', 
units of the package. 

The worldline mass operators are currently extracted up to the 3PN order. Yet, the Feynman 
rules derived from the bulk action, and worldline mass operators of the effective action, in 
eqs.~\eqref{seffsingobj}, \eqref{Spp}, can be currently obtained to any desirable PN order. 
Hence this unit is sufficient in order to compute higher PN order point mass sectors. The 
worldline spin operators are extracted up to the NNLO in the quadratic order in the spin, 
and up to the LO in the cubic and quartic order in the spin, which eventually contribute up 
to the 4PN order in the two-body interaction.

\subsection{FeynGen: Generating Feynman graphs}
\label{feyngen}

``FeynGen'' is the unit, which generates the graph topologies up to a certain power in $G$, 
Newton's $d$-dimensional gravitational constant, and the Feynman diagrams up to a certain PN 
order, in the EFT of PNG for the binary including spins.

Our approach is to consider a ``vertices'' manifold with the propagator (or two-point 
function), as the metric, and the $n$-point functions (or vertices) as tensors of rank $n$. 
Hence, the Wick contractions are actually handled as metric contractions, such that the 
spacetime points (including worldline points) are indices on the related tangent bundle. The 
code uses basic combinatoric and graph features of Mathematica, together with the power of 
abstract tensor algebra that xTensor provides, to automatically create all possible graph 
contractions, including each of the related symmetry factors. 

For a $G^\text{n}$ topology one should consider in general up to $n$-graviton worldline 
vertices, and up to $(n+1)$-point functions. The topologies are then filtered to 
satisfy the following requirements: 1.~They should contain two worldlines; 2.~They should  
be tree level graphs (worldlines are stripped); 3.~They should \textit{not} be UV 
renormalization topologies, i.e.~such that renormalize the Wilson coefficients of the EFT, 
i.e.~the mass, spin, or spin-induced multipole coefficients; 4.~They should be connected 
graphs (again - worldlines stripped). The code generates the graphs corresponding to all 
possible contractions.    

The output of ``FeynRul'' is then imported, and the topological vertices and propagators are 
assigned with the specific NRG fields in the Feynman rules. In addition, up to $n$ 
propagator time insertions are added separately, to the nPN order. All in all, this 
generates all the PN weighted contractions/graphs, and the diagrams can be projected to the 
various point mass and spin sectors according to their worldline vertices. An output of the 
contractions, and of their graphic visualization, is produced both for the topologies, and 
the PN diagrams. The run time of this unit is very short (of the order of minutes).

\subsection{NLoop: Producing loop master integrals} 
\label{nloop}

The ``NLoop'' unit produces all the loop master integrals required for the ``Main'' unit to 
perform the Feynman integration. We recall that all the integrals are evaluated using 
dimensional regularization in a generic dimension $d$.

The unit begins with the Fourier integrals required at the zero-loop level, continues with 
the one-loop integrals, and currently also includes performing all the integration by parts 
(IBP) reductions, required to disentangle the irreducible two-loop integrals. For the other 
simpler two-loop integrals the Fourier and one-loop integrals produced in this unit, 
together with the various integral manipulations carried out in the ``FeynInteg'' section of 
the ``Main'' unit are sufficient. As we noted in section \ref{overview}, when using the NRG 
fields, the two-loop level suffices to handle the 3PN order in the point mass sector, and up 
to the NNLO, up to quadratic in the spins, that is up to the 4PN order with spins.

For both the Fourier and the one-loop integrals, we start with the scalar master integrals 
\cite{Levi:2011eq}, and derive iteratively the required tensor integrals up to rank 6. 
For the irreducible two-loop integrals, we make IBP reductions, and we further process and 
simplify the resulting two-loop integrals, using the lower order loop integrals, in order to 
avoid repeated evaluations each time an irreducible two-loop integral is encountered, when 
the diagrams are automatically evaluated in the ``FeynComp'' section of the ``Main'' unit. 
We produce the IBP reductions of the irreducible two-loop integrals from the scalar 
integral to the rank 6 tensor ones, where the tensor indices can belong to each of the two 
loop momenta, and the derived reductions are mirrored to cover all symmetric cases.

The treatment of the irreducible two-loop integrals is currently substantially more
time-consuming, compared to e.g.~the quick ``FeynRul'' or ``FeynGen'' units. The overall run 
time of the ``NLoop'' unit is currently of the order of 6 hours. Yet this unit should only 
be run once, in order to generate the output file of loop master integrals exported to the 
``Main'' unit, independent of which sectors and diagrams are eventually chosen to be 
evaluated. We expect to extend this unit to handle higher loop levels in the future.

\section{The gauge invariant observables pipeline}
\label{giobs}

Eventually, we get from the ``Main'' unit the two-body effective action, in particular the 
two-body interaction potential, obtained from the Feynman diagrams in the EFT of PNG of the 
binary including spins. From this point on there is no need to consider the EFT of PNG in 
the analysis any longer. This is since the corresponding Hamiltonians and physical EOMs are 
derived in a straightforward manner from the resulting effective actions that our EFT
formulation provides, as described in \cite{Levi:2014sba,Levi:2015msa}. 

In the final ``GI observables'' unit we provide the pipeline to obtain possible 
derivatives and GI observables of interest. In addition to the physical EOMs and the 
Hamiltonians, which are essential for the EOB models of the gravitational waveforms, we also 
derive the conserved global integrals of motion, and the binding energies, as presented, 
e.g., in \cite{Levi:2014sba,Levi:2015uxa,Levi:2016ofk}. We expect this unit in particular to 
be extended by the community, including the authors, to go as far as to include the GW waveform and flux.

\section{Conclusions} 
\label{theendmyfriend}

In this paper we presented a novel public package ``EFTofPNG'', which we created for high 
precision computation in the EFT of PN Gravity, including spins. We created this package in 
view of the timely need to publicly share automated computation tools, which could then be 
assembled to even more comprehensive public codes, that integrate the various types of 
physics manifested in the expected increasing influx of GW data. In particular the inspiral 
phase of the binary, which consists in general spinning components, can only be described 
analytically via the PN theory of GR. Moreover, the EFT approach is naturally tailored for 
high precision computation. Hence, we aimed to make the EFT of PNG, including spins 
\cite{Goldberger:2004jt,Goldberger:2007hy,Levi:2015msa}, accessible in its most practical 
form to the classical Gravity community. 

Our goal was then to create a free and open source package, which is self-contained, 
modular, all-inclusive, and accessible to the Gravity community. Indeed, the ``EFTofPNG'' 
package is written in Mathematica, and also uses the power of the ``xTensor'' package, 
naturally suited for the complicate tensor computation required in Gravity, and which is 
familiar to the Gravity community. Moreover, our coding strategically approaches the generic 
generation of Feynman contractions and graphs, which is universal to all perturbation 
theories in physics, by efficiently treating $n$-point functions as tensors of rank $n$, 
thus again benefiting from the power of ``xTensor'', while also being of potential use for a 
broader community of Physics.

The package currently contains four independent units, which serve as subsidiaries 
to the main one: 1.~The ``FeynRul'' unit, which evaluates the Feynman rules; 2.~The 
``FeynGen'' unit, that generates the required Feynman contractions and diagrams; 3.~The 
``NLoop'' unit, which produces all the required loop master integrals; 4.~The ``Main'' unit, 
which performs the complete evaluation of the Feynman diagrams; 5.~The final ``GI 
observables'' unit, which provides the full computation of derivatives and gauge 
invariant physical observables of interest, and serves as a pipeline chain for the 
obtainment of the final GW templates for the detectors. 

The upcoming ``EFTofPNG'' package version 1.0 should cover the point mass sector up 
to the 3PN order, and all the spin sectors up to the 4PN order, including the NNLO up to 
quadratic in the spins 
\cite{Levi:2011eq,Levi:2014sba,Levi:2015uxa,Levi:2015ixa,Levi:2016ofk}, and the LO cubic and 
quartic in the spins \cite{Levi:2014gsa}, based on the formulation of EFT including spins in 
\cite{Levi:2015msa}. The current preliminary version 0.99 is already released as a public 
repository in GitHub, and can be found in the URL: 
``https://github.com/miche-levi/pncbc-eftofpng''. We expect and strongly encourage public 
development of the package to improve its efficiency, and to extend it to further PN 
sectors, and observables useful for the waveform modelling, as we enter 
the high precision Gravity era.

\acknowledgments

ML is grateful to John Joseph Carrasco for his continuous encouragement and support.
ML would also like to thank Guillaume Faye for pleasant discussions on xTensor.
The work of ML is supported by the European Research Council
under the European Union's Horizon 2020 Framework Programme 
FP8/2014-2020 grant no.~639729, preQFT project.

\bibliographystyle{jhep}
\bibliography{gwbibtex}

\end{document}